\documentclass[twocolumn,showpacs,preprintnumbers,amsmath,amssymb]{revtex4}

\usepackage{graphicx}
\usepackage{dcolumn}
\usepackage{bm}

\begin{document}
\title{Thermodynamic interpretation of the origin of the Universe.}
\author{B. I.~Lev$^{1,2}$}
\author{A.G. Zagorodny$^{1}$}
\affiliation{$^{1}$Bogolyubov Institute for Theoretical Physics, National Academy of Science, 
Metrolohichna St. 14-b, Kyiv 03680, Ukraine,
$^{2}$Department of Physics, Chungnam National University,220 Gung-dong, Yuseong-gu, 
Daejeon 305-764, Korea}
\date{\today}

\begin{abstract}
We employ the law of increasing entropy and the assumption about the decrease of the ground state energy to answer the question 
what are the reasons of the Big Bang and the origin of the Universe.
\end{abstract}
\maketitle

Within the context of the fundamental principles of thermodynamics, any macroscopic system embedded in a thermal bath 
approaches equilibrium  during some relaxation time. In  the equilibrium state, the properties of the system do not 
depend on the way of how the equilibrium has been established. The equilibrium state is, however, realized only under 
certain idealized conditions, so in  reality the properties of the system in a quasi-stationary (steady) state can  
depend both on the  specifics of the  interaction  of the system with the thermal bath and the characteristics of 
the bath \cite{Huang}- \cite{Lan}. The same concerns nonequilibrium systems.
 
Our Universe is non-equilibrium \textsl{a priory}. Therefore, to study its evolution, we  have to choose  a method 
for the description of non-equilibrium systems. Although such systems are widely present in the nature and laboratories, 
there are no unified commonly accepted theoretical approaches  that would make it possible to determine  probable 
states of quasi-stationary nonequilibrium systems. Hence, it is a fundamental task to develop a method for exploring 
the general properties of steady states of open systems and to  reveal the conditions  for such states to exist. 
One of possible ways  to solve this  general problem can be based on Gibbs'es approach \cite{Gibbs}. 
The main  purpose of  our latter is to propose a simple way of describing non-equilibrium systems in the energy  
space \cite{Lev1}, \cite{Lev2} and to formulate a new concept of the solution of the cosmological problem. 
We start from the formulation of the statistical approach.

The canonical partition function in the phase space in the equilibrium case can be written in the form:
\begin{equation}
\rho(q,p)d\Gamma=\exp\{\frac{F-H(q,p)}{\Theta}\} d\Gamma
\end{equation}
where $H(q,p)=E $ is the Hamiltonian on the hypersurface of  constant energy $E$, $d\Gamma=\prod_{i}dq_{i}dp_{i}$ 
is an element of the phase space, $\Theta=kT$, $T$ is the temperature, and $F$ is the free energy that can be  found from the
normalization condition $ \int \exp(\frac{F-H(q,p)}{\Theta})d\Gamma=1.$. As is shown in \cite{Gibbs}, the phase space depends
only  on energy and external parameters.  We introduce an additional function $\Sigma=\ln\frac{d\Gamma}{d E} $, then 
the canonical partition function reduces to the form given by
\begin{equation}
\rho(E)dE=C \exp\{\frac{F-E}{\Theta}+\Sigma(E)\}dE. \label{11}
\end{equation}
The latter presentation makes it possible to describe the dependence of the distribution function  on the energy of the 
macroscopic system \cite{Gibbs}. The normalization condition in this presentation can be written in the form
 \begin{equation}
 \int c \exp(\frac{F-E}{\Theta}+\Sigma(E)) dE=1.
\end{equation}
from which one can find the normalization constant which take into account the determinant transformation between phase space
and energy variable. In order to select the states  that give the dominant contribution to the partition function  we employ 
the condition $\frac{d\Sigma}{dE}=\frac{1}{\Theta}$ that determines the temperature of the system provided the change of the 
phase space as a function of the system energy is known. Using this definition and taking into account the basic principles 
of statistical mechanics \cite{Lan} we come to the conclusion that $\Sigma=\ln\frac{d\Gamma}{d E}= S $ is equal to the entropy 
of the system.  Now we have to make a very important notice: the temperature describes the dependence of the entropy only on 
energy, but not on any other thermodynamic quantities. We can define the temperature for other situations, but this definition 
has no sense without changing entropy.  Another important conclusion is that we can calculate the partition function by integration 
over energy. Such integration in this sense means the continual integral on  the energy variable. The extremum of the partition function 
is realized  under the condition: $F=E-\theta S$ and  any probable deviation from this condition gives very small contribution 
to macroscopic characteristics similar to the quantum contribution to classical trajectories. The novelty of this 
presentation consists in the possibility to use the kinetic equations for the description of non-equilibrium systems as 
Brownian systems in the energy space \cite{Lev2}. Starting from the basic kinetic equations for the distribution 
function of the macroscopic system in the energy space, we can obtain steady states and fluctuation dissipation relations 
for non-equilibrium systems \cite{Lev1}. 

We introduce an additional internal ``time'' parameter  and thus obtain a very important relation
\begin{equation}
\dot{S}=\frac{dS}{dE}\dot{E}=\frac{1}{\Theta}\dot{E}
\end{equation}
that means that  entropy changes with time determine energy changes as a function of the internal parameter. 
Here and in what follows $\dot{E}$ denotes the derivative  with respect to this internal ``time'' parameter.  
Within the context of the law of increasing entropy, the thermodynamic relation $\dot{S}>0$ and the condition 
of the phase space increase with energy  changes,  we find that $\dot{E}>0$ if the system relaxes to the equilibrium 
state as the state with maximum entropy. We can determine the internal parameter provided  energy changes and 
the system entropy are known. The internal parameter ``time'' determines the increase of entropy with energy increase. 
On the other hand, the evolution of the system leads to the state with the minimum energy. Thus the question arises 
whether there exist  processes with positive changes of energy in time  under decreasing energy of the system. 

In this paradigm it is possible to introduce an important general relation between pressure and concentration in vacuum state.  
Let us assume that the initial vacuum state has energy $E_{v}$.  By virtue of quantum nature of  vacuum in such initial state, 
fluctuations of all  probable fields are generated. The first problem is to formulate the equation of state for vacuum. The 
thermodynamic relation yields the pressure of vacuum $P=-\Theta \frac{d S}{dV}$ where $V$ is the volume. So
\begin{equation}
P==-\Theta \frac{d S}{d E_{v}}\frac{d E_{v}}{d V} =-\frac{d E_{v}}{dV}=-\rho_{v}
\end{equation}
where  the additive energy $E_{v}=\rho_{v}V$ is introduced, with $\rho_{v}$ being the energy density. It is the general 
equation of state for vacuum.  It follows from the statistical principles, for the changes of the probable states of the 
system that accompany the entropy increase.  We can suggest that this is the general dynamical principle for all the states 
of macroscopic systems. 

In thermodynamics heat can be treated as the energy that flows between the degrees of freedom  cannot be macroscopically observable. 
We define it as $dQ=\Theta dS$. According to Boltzmann \cite{Boltzmann}, and if we can heat a macro-system, that means that this system 
has micro-structure. The primary vacuum state is the only state which has not structure, but it is possible to heat this state after field 
or particle origin. That satisfies the third low of thermodynamics. The temperature of the vacuum state is equal to zero and it does not depend on the system 
energy.  We begin to change the states of vacuum  and thus  can determine the heat change as 
\begin{equation}
\dot{Q}=\Theta \dot{S}=\Theta \frac{dS}{dE}\dot{E}=\dot{E}.
\end{equation}
It follows from this relation that the change of heat determines the energy change and introduces the internal parameter.  Naturally, 
heating will be possible only due to the change of the system energy  $\frac{dE}{dt}>0$ and the system relaxes to the equilibrium state. 
In the general case, for different states of the system, we can take into account the thermodynamic relation 
\begin{equation}
dE+PdV=\Theta S.
\end{equation}
if the heat flow is absent. Then we substitute into this relation the  quantities $V=V_{0}a^{3}$ and $E=V_{0}a^{3}\rho$ where $V_{0}$
are accompany volume considered area and $a=a(t)$ is the causes scales factor in cosmology we can calculate the time derivative of the 
thermodynamic relation  and obtain the equation given by
\begin{equation}
\dot{\rho}+3H(\rho +P)=0
\end{equation}
For the vacuum state $\rho=\rho_{v}$, then  $\dot{\rho_{v}}=0$, i.e., $\rho_{v}=const $ everywhere. 
Usually this thermodynamic equation is used to describe the density evolution of adiabatic processes,  especially in modern 
cosmological evolution models for the Universe.  

To describe the Universe dynamics,  we have to assume that fields generated after the Universe origin  possess certain properties, 
namely they lead to the formation of  new phases that change the energy of the system and induce the entropy increase. 
So, we have to suggest the possibility of decrease of the energy of vacuum state. It does not violate the physical laws since 
the generation of new fields changes the energy of the previous state and  produces a new state of the vacuum. The entropy of 
the initial state is completely determined and takes zero value. It is reasonable to assume that some  portion of the energy of 
the generated fields  spent for the stabilization of states of the new fields  within the context of the nonlinear field dynamics. 
Let us consider the arguments in favor of such possibility. 

Standard cosmological models involve a scenario of the Universe nucleation and expansion based on a scalar field  that is of 
fundamental importance for the unified theories of weak, strong, and electromagnetic interactions with 
spontaneous symmetry breaking. What  can we say about the behavior of the fundamental scalar field?. 
Let us assume that the scalar field generated in vacuum can increase the energy and we can write an expression for the new energy as given by
\begin{equation}
E=E_{v}+V(\varphi)
\end{equation}
where the second term  is associated with the energy of the scalar field. We consider the situation when the new states have energies 
smaller than the energy of the initial state of vacuum. In this case we can write 
\begin{equation}
E=E_{v}-\frac{\mu ^{2}}{2}\varphi ^{2}
\end{equation}
where the second term takes into account the symmetry properties of the scalar field energy. As shown below $V(\varphi)$ is negative 
that means the conservation of energy take place. The coefficient $\mu ^{2}$ describes the coupling of the ground state with the new 
fields. This coefficient cannot be constant by virtue of  probable fluctuations related to other fields which can be generated in 
vacuum. We can take into account these fluctuations in the following way. The partition function for the system under consideration 
can be written in the standard form
\begin{equation}
Z\sim \int D \varphi \int D\xi \exp \frac{1}{\Theta}\left\{ -E_{v}+\frac 12\mu ^2\varphi ^2-\frac 12 \xi \varphi^2 -\frac{\xi ^2}{\sigma ^2}\right\}
\end{equation}
where the coefficient $\mu ^2=\mu ^2+\xi $ is presented as a sum of the average value and the fluctuation part  that stochastically
changes this value in the course of interactions of the scalar field with  probable fluctuations of  different nature.  The last part 
in the exponent  is associated with the energy of  fluctuations of  different nature with disperse $\sigma ^2$.  We perform continuous 
integration over all  probable fluctuation $\xi$ and  make use of the property of the Gauss integral. Thus we obtain  a relation for 
the partition function,
\begin{equation}
Z\sim \int D \varphi \int D\xi \exp \frac{1}{\Theta}\left\{ -E_{v}+\frac 12\mu ^2\varphi ^2 -\frac{\sigma ^2 \varphi ^4}{4}\right\}
\end{equation},
that determines the new energy 
\begin{equation}
E=E_{v}-\frac 12\mu ^2\varphi ^2+\frac{\sigma ^2 \varphi ^4}{4}
\end{equation}
where $V(\varphi)=-\frac 12\mu ^2\varphi ^2+\frac{\sigma ^2 \varphi ^4}{4}$ is the well-known standard energy of the fundamental scalar 
field. The total energy of the new ground state and the fundamental scalar field can be presented as 
\begin{equation}
E=E_{v}-\frac{\mu ^4}{4 \sigma^{2}}+\frac{\sigma\mu ^2}{4}(\varphi^2-\frac{\mu ^2}{\sigma^{2}})^{2}
\end{equation}
with the following asymptotic behavior: if $\varphi=0$, then we have the ground state energy $E=E_{v}$,; if $\varphi^2=\frac{\mu ^2}{\sigma^{2}}$, 
then the energy takes the value $E=E_{v}-\frac{\mu ^4}{4 \sigma^{2}}$. This energy can  vanish for vanishing $\sigma^{2}$ (dispersion
of fluctuations of the fields of  different nature) . If $\sigma^{2}$  tends to infinity, then the energy of the new state tends to the initial 
energy of the ground state. The energy of the new state can  vanish for $E_{v}=\frac{\mu ^4}{4 \sigma^{2}}$. This relation can be applied to estimate
the maximum dispersion of the field fluctuations if the vacuum  temperature is given by $\Theta_{v}$ 
$\sigma^{2}=\frac{\mu ^4}{2 \Theta_{v}}$. Thus, we can point out that we  have come to the standard form of the energy of the fundamental scalar 
field, but  with different behavior of the energy of vacuum at the presence of the scalar field. The coefficient of nonlinearity in the potential 
energy is determined by the coupling of the fundamental scalar field with the fluctuations of the field of  different nature. This means that 
there could be a new scenario of the Universe formation.

In this case we can write the time derivative of the energy of new states as 
\begin{equation}
\dot{E}=\dot{E_{v}}+\frac{dV}{d \varphi}\dot{\varphi}.
\end{equation}
On the other hand, from the cosmological models we know that for slowly changing homogeneous fundamental scalar field \cite{LIN}
\begin{equation}
\frac{d \varphi}{dt}=-\frac{1}{3H}\frac{dV}{d \varphi} 
\end{equation}
where $H$ is the Hubble constant. In this case all necessary conditions satisfy the thermodynamic relation 
\begin{equation}
\dot{E}=\dot{E_{v}}-\frac{1}{3H}(\frac{dV}{d \varphi})^{2}>0
\end{equation}
or 
\begin{equation}
\dot{E_{v}}>\frac{1}{3H}(\frac{dV}{d \varphi})^{2}
\end{equation}
 that can realize the increasing entropy of the new states. Substituting the thermodynamic relation for the vacuum state $\dot{E_{v}}-3\rho_{v}a^{2}\dot{a}=0$ 
or $\dot{E_{v}}=3E_{v}H$ we can obtain the necessary condition of the Universe formation with nonzero fundamental scalar field, i.e.,
\begin{equation}
E_{v}>(\frac{dV}{3H d \varphi})^{2}>\dot{\varphi}^{2}=2T
\end{equation}. 
This relation implies that the vacuum energy should be larger than  twice the kinetic energy $T$ of the additional fundamental scalar field. 
We have the entropy increase  with time  and decrease of the energy of the new state  that is smaller than the energy of the initial ground 
state (vacuum). To some extent such possibility means that the system in new state is heated that could be  employed as an interpretation of 
the Big Bang! The adiabatic expansion of the Universe is possible, if the doubled kinetic energy of the scalar field is equal to the initial 
fluctuation energy of the field of different nature. 
 
All the cosmological models start from the Big Bang  that generates the heated Universe. The present approach can change the starting point. 
We can start from the vacuum state with  fluctuations of various nature. At the starting ``time'' when the additional scalar field appears 
the number of possible states increases and the energy of the new vacuum state decreases due to the formation of the growing bubbles of the new state. 
Within this interval the standard cosmological models work. In particular, for the critical temperature \cite{LIN}
$\Theta_{c}=\frac{2 \mu }{\sigma}$, the symmetry in the behavior of the fundamental scalar field restores and we have the usual well-known 
behavior of the Universe, but with principal difference. The initial vacuum is not completely restored, and we can observe the 
expansion of the Universe with acceleration. As was mentioned above it depends on the dispersion of fluctuations of the fields of  different 
nature, but in the case of the new phase bubble formation not all possible fluctuations can influence the processes inside the Universe. This 
effect is known as effect of fluctuation condensation \cite{Lev3}. In such a case the acceleration never stops.

To describe the advantages of the presentation of the canonical partition function discussed above, we can present a recent approach to the 
description of non-equilibrium macroscopic systems \cite{Lev1},\cite{Lev2}. Such approaches with experimental accessibility of the generalized 
fluctuation-dissipation relations for the non-equilibrium steady states have been used for the description of a non-equilibrium system. The solutions 
of this problem require the modified fluctuation-dissipation and $E $ Einstein relations for non-equilibrium steady states or exactly solvable 
canonical models.  A natural hypothesis arises, that the non-equilibrium distribution functionl depends on the energy of the system \cite{Gibbs}. 
In a similar way, our considerations are based on the energy representation, where the states of the system are determined  only by their energies. 
The variation of energy can induce the variation of the states of the macroscopic system. In the general case the non-equilibrium system 
can change the state with some energy value to another one and this process depends on the external  influence and the initial conditions. 
The external effects are manifested first of all in the variation of the system energy. The energy is also  changed if the system looses, or 
absorbs energy  due to the initial value \cite{Lev1}. 

As was shown before, the stable vacuum states depend on the nature and properties of fluctuations in the system under consideration. Specifically, 
they are influenced by all  probable multiplicative fluctuations which in turn determine the probabilities of transitions between stable vacuum states.  
To calculate such transition probabilities, we should consider the interaction with the multiplicative noise regarding possible changes of both 
dispersion of  fluctuations and the potential of the scalar field. In this case the non-linearity of the scalar field potential and fluctuations 
of the field of  different nature should be taken into account. Such fluctuations can change the minimum of the potential and determine the 
alternative way of the Universe formation. For example, in  this case the state of the Universe can be characterized by $\varphi=0$ but with 
nonzero Hubble constant. Thus, we have proposed  a model for  the description of the non-equilibrium Universe  that makes it possible to determine 
new stationary states. The stationary distribution functions of the Universe have been obtained for the standard formation but regarding the 
interaction with the nonlinear environment. The presented treatment assumes the presence of  a negative mass coefficient in the equation for 
the potential of the fundamental scalar field. The presence of the negative mass coefficient is responsible for the unstable ground state and 
the appearance of the fluctuations  that can be taken into account only in this approach.

Acknowledgment: This article carry out for financial support from theme department of physics and astronomy of NAS Ukraine 
«Dynamic formation spatial uniform structures in many-body system » PK 0118U003535. This work was supported in part by Brain 
Pool programm by Grant N 218H1D3A2065894 through the National Research Foundation of Korea (NRF).

\end{document}